\documentclass[12pt,a4paper,dvips]{article}
\usepackage{a4p}
\usepackage{cite,mcite}
\usepackage{graphicx}
\usepackage{physics}
\usepackage{l3_title}
\usepackage{rotating}
\journalname{Phys. Lett. B}
\date{July 19, 2000}
\preprint{2000-097}
\newlength{\capindent}
\newlength{\capwidth}
\newlength{\figwidth}
\newcommand{\icaption}[2][!*!,!]{\hspace*{\capindent}%
  \begin{minipage}{\capwidth}
    \ifthenelse{\equal{#1}{!*!,!}}%
      {\caption{#2}}%
      {\caption[#1]{#2}}
  \end{minipage}}
\def \be {\begin{equation}}
\def \e {\end{equation}}
\def \bea {\begin{eqnarray}}
\def \ea {\end{eqnarray}}

\def \g {\gamma}
\def \ao {$a_0/\Lambda^2$}
\def \ac {$a_c/\Lambda^2$}

\def \wwg {$\Wp\Wm\gamma$ }
\def \nngg {$\nu\bar\nu\gamma\gamma$ }
\def \xs {cross section }

\def \no {\nonumber}

\newcommand{\To}[2]{\stackrel{#1}{\hbox to #2 pt{\rightarrowfill}}}

\def \vector#1{\stackrel{\hspace{-0.45em}\longrightarrow}{#1}}

\begin{document}

\begin{titlepage}
  \title{Measurement of the \boldmath{$\Wp\Wm\gamma$} Cross Section \\
    and Direct Limits on Anomalous Quartic\\
    Gauge Boson Couplings at LEP\\}
  
  \author{The L3 Collaboration}

%
%
\begin{abstract}
  
  The process $\epem\ra\Wp\Wm\gamma$ is analysed using the data
  collected with the L3 detector at LEP at a centre-of-mass energy of
  $188.6\GeV$, corresponding to an integrated luminosity of
  176.8\,pb$^{-1}$.  Based on a sample of 42 selected $\Wp\Wm$
  candidates containing an isolated hard photon, the $\Wp\Wm\gamma$
  cross section, defined within phase-space cuts, is measured to be:
  $\sigma_{{\rm WW}\gamma} = 290 \pm 80 \pm 16$ fb, consistent with
  the Standard Model expectation.
  
  Including the process $\epem \ra \nu\bar\nu \g\g$, limits are
  derived on anomalous contributions to the Standard Model quartic
  vertices $\Wp\Wm\gamma\gamma$ and $\Wp\Wm\Zo\gamma$ at 95\%~CL:
\begin{eqnarray*}  
-0.043  \GeV^{-2} & < & a_0/\Lambda^2 ~ < ~ 0.043 \GeV^{-2} \\ 
\phantom{0}-0.08 \GeV^{-2} & < & a_c/\Lambda^2 ~ < ~ 0.13\phantom{0} \GeV^{-2} \\ 
\phantom{0}-0.41 \GeV^{-2} & < & a_n/\Lambda^2 ~ < ~ 0.37\phantom{0} \GeV^{-2}.
\end{eqnarray*}
\end{abstract}

\vspace*{15mm} \centerline{Submitted to {\it Phys. Lett. B } }

\end{titlepage}

\section*{Introduction}

\begin{figure}[htbp]
\begin{center}
  \mbox{\normalsize
\begin{picture}(15000,35000)(-15000,-42000)
\THICKLINES    
\drawline\photon[\E\REG](-15000,0)[7]
\global\advance\pmidx by -1000
\global\advance\pmidy by  1000
\put(\pmidx,\pmidy){{\Large $\gamma/$Z}}
\drawline\photon[\NE\REG](\photonbackx,\photonbacky)[6]
\global\advance\photonbackx by 500
\put(\photonbackx,\photonbacky){{\Large W$^+$}}
\drawline\photon[\SE\REG](\photonfrontx,\photonfronty)[6]
\global\advance\photonbackx by 500
\global\advance\photonbacky by -1500
\put(\photonbackx,\photonbacky){{\Large W$^-$}}
\drawline\photon[\E\REG](\photonfrontx,\photonfronty)[6]
\global\advance\pmidx by +1600
\global\advance\pmidy by +1000
\put(\pmidx,\pmidy){{\Large $\gamma$}}
{\put(\photonfrontx,\photonfronty){\circle*{500}}}
\global\advance\pmidx by -1000
\drawline\fermion[\SW\REG](-15000,0)[5500]
\global\advance\pmidx by -4000
\global\advance\pmidy by -3200
\put(\pmidx,\pmidy){{\Large e$^-$}}
\drawline\fermion[\NW\REG](-15000,0)[5500]
\global\advance\pmidx by -3800
\global\advance\pmidy by +2000
\put(\pmidx,\pmidy){{\Large e$^+$}}
\global\advance\pmidx by -4800
\global\advance\pmidy by +2000
\put(\pmidx,\pmidy){\Large a)}
\global\advance\pmidy by -16000
\put(\pmidx,\pmidy){\Large b)}
\drawline\fermion[\NW\REG](-11000,-18000)[8000]
\global\advance\pmidx by -4000
\global\advance\pmidy by +3200
\put(\pmidx,\pmidy){{\Large e$^+$}}
\drawline\fermion[\NE\REG](-11000,-18000)[8000]
\global\advance\pmidx by 3100
\global\advance\pmidy by 3200
\put(\pmidx,\pmidy){{\Large $\bar{\nu_{\mathrm{e}}}$}}
\drawline\photon[\S\REG](-11000,-18000)[4]
\global\advance\pmidx by -3200
\global\advance\pmidy by -850
\put(\pmidx,\pmidy){{\Large W$^+$}}
\drawline\photon[\S\REG](\photonbackx,\photonbacky)[4]
\global\advance\pmidx by -3200
\global\advance\pmidy by -850
\put(\pmidx,\pmidy){{\Large W$^-$}}
\drawline\fermion[\SW\REG](\photonbackx,\photonbacky)[8000]
\global\advance\pmidx by -4000
\global\advance\pmidy by -4250
\put(\pmidx,\pmidy){{\Large e$^-$}}
\drawline\fermion[\SE\REG](\photonbackx,\photonbacky)[8000]
\global\advance\pmidx by 3100
\global\advance\pmidy by -4050
\put(\pmidx,\pmidy){{\Large ${\nu_{\mathrm{e}}}$}}
\drawline\photon[\NE\REG](\photonfrontx,\photonfronty)[8]
\global\advance\photonbackx by 500
\put(\photonbackx,\photonbacky){{\Large $\gamma$}}
\drawline\photon[\SE\REG](\photonfrontx,\photonfronty)[8]
\global\advance\photonbackx by 500
\global\advance\photonbacky by -500
\put(\photonbackx,\photonbacky){{\Large $\gamma$}}
{\put(\photonfrontx,\photonfronty){\circle*{500}}}
\end{picture}
}
\caption{Feynman diagrams containing a four-boson
  vertex leading to the (a) \wwg and to the (b) 
  $\nu_{\mathrm{e}}\bar\nu_{\mathrm{e}} \g\g$ final states.}
\label{fey}
\end{center}
\end{figure}
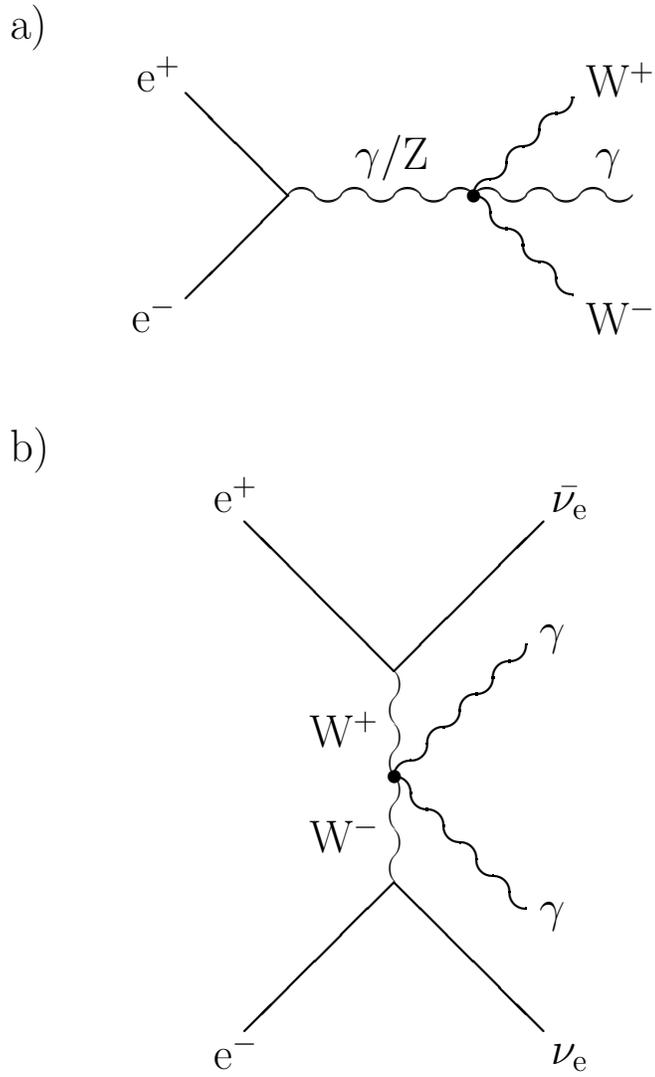

The LEP centre-of-mass energy for $\epem$ collisions is now well above
the kinematic threshold for W-pair production allowing for the study
of radiative W-pair production, $\epem\ra\Wp\Wm\gamma$.  The Standard
Model (SM)~\cite{sm_glashow,velt} predicts the
existence of quartic gauge couplings (QGCs), leading to $\Wp\Wm\gamma$
production via $s$-channel exchange of a $\gamma$ or Z boson as shown
in Figure~\ref{fey}a.

As the contribution of these two quartic Feynman diagrams with respect
to the other competing diagrams, mainly initial-state radiation, is
negligible at the LEP centre-of-mass energies, the process leading to
the \wwg final state could thus be sensitive to anomalous
contributions to the SM quartic gauge-boson vertices
$\Wp\Wm\gamma\gamma$ and $\Wp\Wm\Zo\gamma$.

The existence of Anomalous QGCs (AQGCs) would also affect the $\epem
\ra \nu_{\mathrm{e}}\bar\nu_{\mathrm{e}} \g\g$ process via the
$\Wp\Wm$ fusion Feynman diagram containing the $\Wp\Wm\gamma\gamma$
vertex~\cite{stir_phot} (see Figure~\ref{fey}b).  In the SM the
reaction $\epem \ra \nu\bar\nu\g\g$ proceeds predominantly through
$s$-channel Z exchange and $t$-channel W exchange, with the two
photons coming from initial state radiation, whereas the SM
contribution from the $\Wp\Wm$ fusion is negligible at LEP.  AQGCs
would enhance the $\nu\bar\nu\g\g$ production rate, especially for the
hard tail of the photon energy distribution and for photons produced
at large angles with respect to the beam direction.

Here we describe the cross section measurement for the process
$\epem\ra\Wp\Wm\gamma$ and the determination of AQGCs using the data
collected in 1998 with the L3 detector~\cite{l3det} at $\sqrt{s}=
188.6\GeV$ (denoted as $\sqrt{s}= 189\GeV$ hereafter) corresponding to
an integrated luminosity of 176.8\,pb$^{-1}$.  AQGCs are also
independently determined using acoplanar photon-pair events with
missing energy.  This analysis is performed using the data at
$\sqrt{s}= 189\GeV$ and at $\sqrt{s}= 182.7\GeV$ collected in~1997
(denoted as $\sqrt{s}= 183\GeV$ hereafter) corresponding to a total
integrated luminosity of 231.7\,pb$^{-1}$.  The results derived on
AQGCs from the \wwg and \nngg channels are finally combined.

The search for anomalous contributions to the SM quartic couplings is
performed within the theoretical framework of
References \citen{bud} and \citen{anja}.  
Recently, experimental measurements for
these couplings have already been performed on final states with three
vector bosons $\Wp\Wm\gamma$~\cite{opal-qgc}
and $\Zo\gamma\gamma$~\cite{zgg}.\\

\section*{\boldmath{$\Wp\Wm\gamma$} Final State and Signal Definition}

There are 14 Feynman diagrams at the tree level leading to the
$\Wp\Wm\gamma$ final state, and many other diagrams corresponding to
photons from the decay products of hadronic or leptonic W's.  
We are interested only in two
of these, the quartic diagrams.  
The other diagrams leading to the same final state are
initial state radiation (ISR), final state radiation (FSR), and
radiation from the W boson itself.

The Monte Carlo used for the \wwg cross section determination is
KORALW~\cite{wk}.  This generator does not include the quartic
coupling diagrams.  Initial state multi-photon radiation is
implemented in KORALW in the full photon phase space.  FSR from
charged leptons in the event up to double bremsstrahlung is included
using the PHOTOS~\cite{photos} package.  Fragmentation processes of
quarks into hadrons are made according to the JETSET~\cite{js}
algorithm including photons in the parton shower. For the
$\Wp\Wm\gamma$ cross section measurement, this modelling is sufficient
since the contribution of all the other diagrams is very small.  The
background processes such as ${\rm e^+e^-}\ra \Zo/\gamma\ra {\rm q
  \bar q}(\gamma)$ and ${\rm e^+e^-}\ra \Zo\Zo\ra 4f(\gamma)$ are
simulated using PYTHIA~\cite{pyt}.  The L3 detector response is
simulated by the program GEANT~\cite{geant}.\\

In this analysis, the $\Wp\Wm\gamma$ signal is defined by the
following phase-space cuts:
\begin{itemize}
\item $E_\gamma >$ 5 GeV, where $E_\gamma$ is the energy of the
  photon,
\item $\theta_\gamma >$ 20$^\circ$, where $\theta_\gamma$ is the angle
  between the photon and the beam axis,
\item $\alpha_\gamma >$ 20$^\circ$, where $\alpha_\gamma$ is the angle
  between the direction of the photon and 
that of the closest charged lepton or jet.
\end{itemize}

These cuts are mainly chosen for experimental reasons, to optimise the
photon identification and the background suppression.
They also largely avoid
any infrared and collinear singularities in the calculation of the
signal cross section.

The theoretically predicted \wwg \xs from KORALW corresponds to
272 $\pm$ 4(stat) fb.  In the EEWWG program~\cite{anja} the effect of
undetected additional ISR collinear to the beam pipe is
included~\cite{priv} by implementing the EXCALIBUR~\cite{exca}
collinear radiator function.  The effect of the higher
order radiative corrections is to move the effective centre-of-mass
energy towards lower values, reducing the expected signal cross
section by about 18\%.  The resulting EEWWG \xs corresponds to
233 $\pm$ 12(theor) fb. This is used as the SM expectation in the
anomalous coupling analysis, which leads to less stringent constraints
on AQGCs.  The theoretical uncertainty~\cite{priv} is propagated to
the AQGC determination.  Consistent results are obtained with the
YFSWW3~\cite{yfs} MC which predicts 224 $\pm$ 6(stat)~fb.
Differences of this order in the predicted \xs with high transverse
momentum photons are expected~\cite{wkpeople} between pure leading-log
and leading-log plus matrix-element based calculations.

\section*{\boldmath{$\Wp\Wm\gamma$} Event Selection and Cross Section}
The W-pair event selections used here are similar to those reported in
Reference \citen{l3-pre98}. Only the semileptonic and fully hadronic
W-pair decay modes are considered.  The number of selected data events
and the expected number of signal and background events are shown in
Table~\ref{tab:ww}.\\

\begin{table}[htb]
\begin{center}
  \renewcommand{\arraystretch}{1.1}
\begin{tabular}{|c||c|c|c|c|}
\hline
Decay Channel & ~~~$N_{\it obs}$~~~ & $\varepsilon_{\rm WW}$ & 
                ~$N^{exp}_{\it TOT}$~ &~$N^{exp}_{\it Bkgr}$~  \\
\hline
\hline
qqe$\nu_e$    & 355  & 0.768$\pm$0.005 & 361 & 19   \\
\hline
qq$\mu\nu_\mu$& 364  & 0.834$\pm$0.002 & 375 & 19   \\
\hline
qq$\tau\nu_\tau$ & 313  & 0.605$\pm$0.003 & 300 & 42 \\
\hline
qqqq           & 1514 & 0.892$\pm$0.006 & 1486 & 296  \\
\hline
\end{tabular}
\caption[]{Number of observed events,
  selection efficiencies with statistical uncertainties, expected
  total number of events and background estimates
  for the various W$^+$W$^-$
  decay channels according to the SM prediction.  The efficiencies
  shown here include the contribution of cross efficiencies from the
  other W-pair decay modes.  }
\label{tab:ww}
\end{center}
\end{table}

\begin{figure}[p]
\begin{center}
  \includegraphics[width=0.95\linewidth]{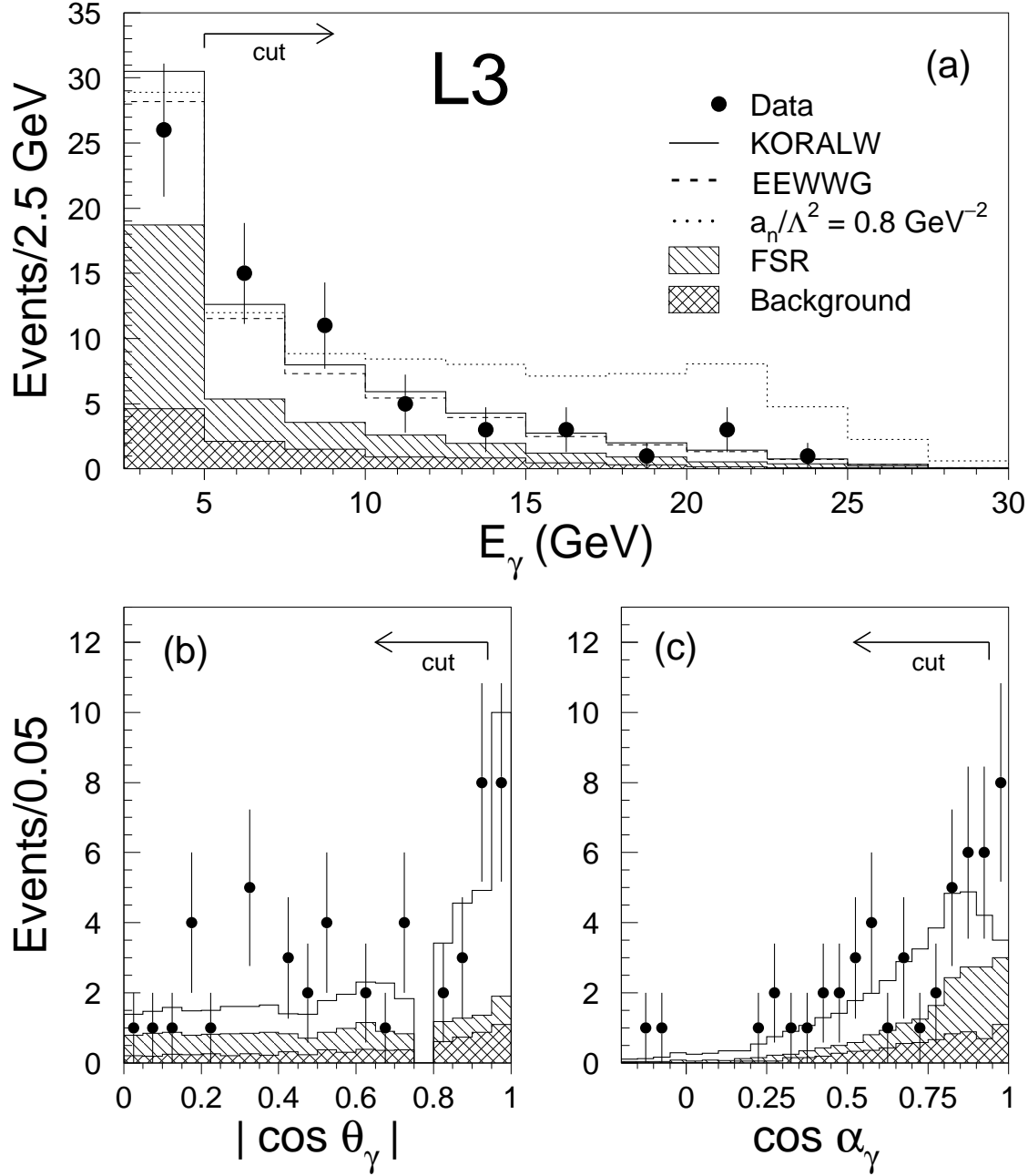}
\caption[]{Differential distributions,
  for the semileptonic and fully hadronic $\Wp\Wm\gamma$ decay modes,
  of (a) the photon energy, (b) the angle of the photon to the beam
  axis and (c) angle of the photon to the closest charged lepton or
  jet.  The hatched area is the background component from ZZ, Zee, and
  ${\rm q \bar q(\gamma)}$ events.  Final state radiation includes the
  contribution of photons radiated off the charged fermions and
  photons originating from isolated meson decays.  In the upper plot,
  the distribution corresponding to a non-zero value of the anomalous
  coupling $a_n/\Lambda^2$ is shown as a dotted line. }
\label{fig:sele}
\end{center}
\end{figure}

The photon selection in $\Wp\Wm$ events is optimised for each
four-fermion final state. Photons are selected by requiring energy
deposition in the electromagnetic calorimeter not associated with any
track in the central detector, and low hadronic activity in a cone of
half-opening angle of 7$^\circ$ around the electromagnetic cluster.
The profile of the shower must be consistent with that of an
electromagnetic particle.  In addition, the highest energy photon
has to satisfy the signal definition requirements:
$E_\gamma>$ 5 \GeV, $\theta_\gamma>$ 20$^\circ$, and $\alpha_\gamma>$
20$^\circ$.  Figure~\ref{fig:sele} shows
the distributions of $E_\gamma$, $\theta_\gamma$, and $\alpha_\gamma$,
where for the last variable the direction of the reconstructed
hadronic jet is assumed as the direction of the quark in the final
state.  In general, good agreement between the data and the SM
expectation is observed.

Table~\ref{tab:wwg} summarises the results for the applied selection
criteria.  In total 42 $\Wp\Wm\gamma$ events are selected, where
37.8$\pm$0.6 is the Monte Carlo expectation.  The efficiency
$\varepsilon_{\rm WW\gamma}$ is defined as the number of selected
KORALW events (regardless of any phase-space cuts) divided by the
number of generated MC events satisfying the signal definition.  It
accounts for small possible migration effects of events from outside
the signal region into the selected sample due to the finite detector
resolution.\\

\begin{table}[htbp]
\begin{center}
  \renewcommand{\arraystretch}{1.1}
\begin{tabular}{|c||c|c|c|c|}
\hline
Decay Channel &~~~$N_{\it obs}$~~~& $\varepsilon_{\rm WW\gamma}$ &~$N^{exp}_{\it TOT}$~&~$N^{exp}_{\it FSR}+N^{exp}_{\it Bkgr}$~\\
\hline
\hline
 qqe$\nu_e\gamma$      & 6  & 0.483$\pm$0.025 & 5.85$\pm$0.26 & 2.26$\pm$0.14 \\
\hline
 qq$\mu\nu_\mu\gamma$  & 5  & 0.547$\pm$0.027 & 6.87$\pm$0.28 & 2.88$\pm$0.17 \\
\hline
 qq$\tau\nu_\tau\gamma$& 7  & 0.351$\pm$0.018 & 4.63$\pm$0.22 & 2.05$\pm$0.14 \\
\hline
 qqqq$\gamma$          & 24 & 0.504$\pm$0.016 &20.4$\pm$0.38 & 9.23$\pm$0.26 \\
\hline
\hline
Total          & 42  & --              &37.8$\pm$0.6 & 16.4$\pm$0.4 \\
\hline
\end{tabular}
\caption[]{Number of observed events, 
  selection efficiencies with statistical uncertainties and expected
  number of total and background events including final state
  radiation.  }
\label{tab:wwg}
\end{center}
\end{table}

The $\Wp\Wm\gamma$ cross section is evaluated channel by channel and
then combined according to the SM
W-pair branching fractions. The result is:
\begin{eqnarray*}
\sigma_{{\rm WW}\gamma} & = & 290 \pm 80 \pm 16 ~{\rm fb}\,,
\end{eqnarray*}
where the first error is statistical and the second systematic.  The
measurement is in good agreement with both the KORALW and EEWWG SM
expectations.

\begin{figure}[p]
\begin{center}
  \includegraphics[width=0.96\linewidth]{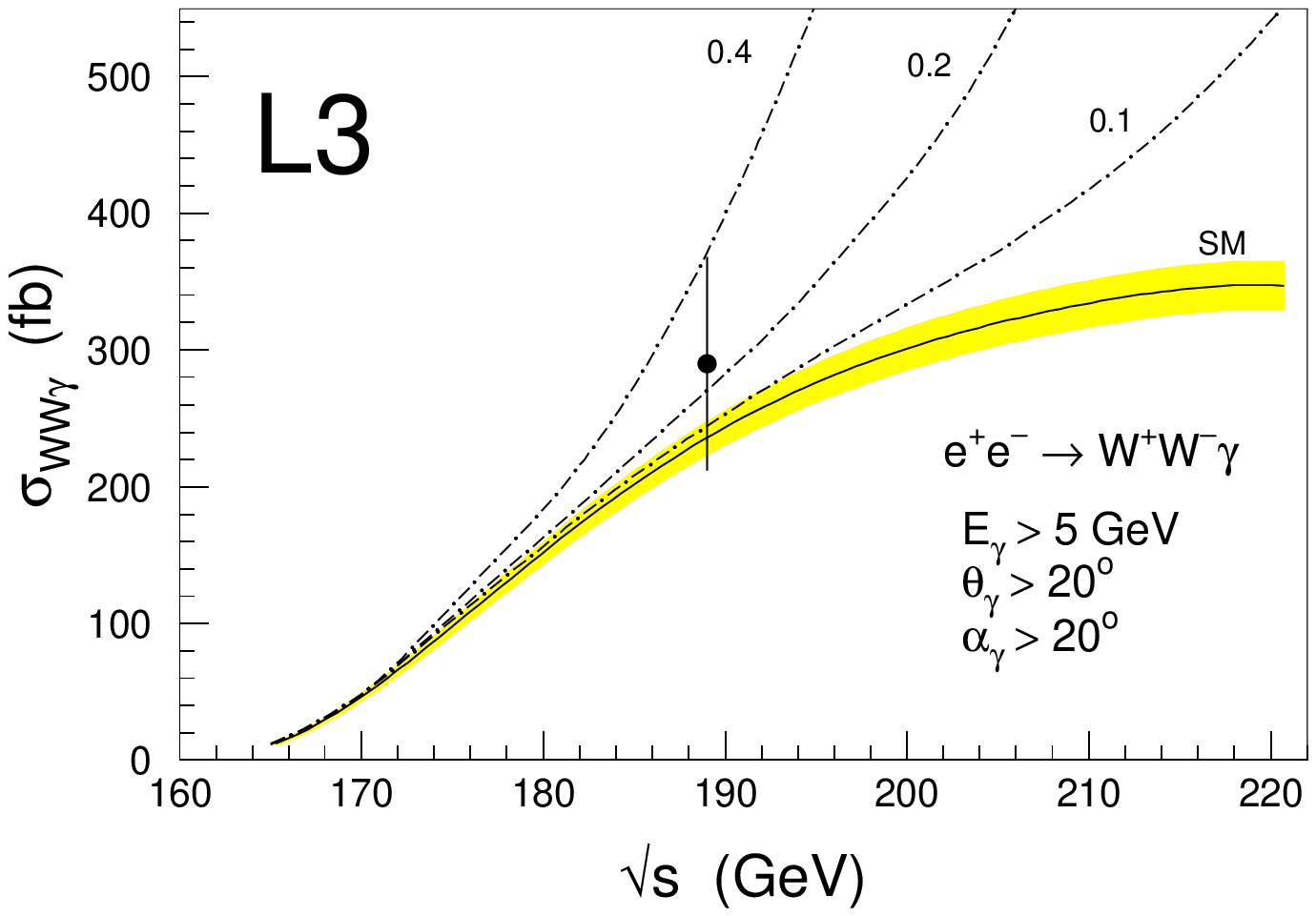}
  \caption[]{
Measured \xs for $\epem\ra\Wp\Wm\gamma$ at $\sqrt{s}=$ 189 GeV (point)
compared to the SM \xs as a function of the centre-of-mass energy
(solid line) as predicted by the EEWWG Monte Carlo within
  the indicated phase-space cuts.  The shaded band corresponds to the
  theoretical uncertainty of $\pm$5\%.  The three dashed lines
  correspond to the \xs for non-vanishing values of the anomalous
  coupling $a_n/\Lambda^2$ (in GeV$^{-2}$ units).}
\label{fig:xsec}
\end{center}
\end{figure}

Figure~\ref{fig:xsec} shows the result obtained together with the
predicted total $\Wp\Wm\gamma$ cross section from the EEWWG Monte
Carlo as a function of the centre-of-mass energy.

The systematic uncertainties arising in the inclusive W-pair event
selections~\cite{l3-pre98} are propagated to the final measurement and
correspond to an uncertainty of $\pm$ 6.3 fb.  Other possible
systematic biases due to detector effects such as electro-magnetic
cluster resolution, angular resolution, and calorimetric energy scale
uncertainty are found to have a negligible effect on the final result.

The total systematic uncertainty is dominated by the JETSET modelling
of photons from meson decays ($\pi^0, \eta$).  To estimate this
effect, a data sample of 3.9 pb$^{-1}$ collected in 1998 at
$\sqrt{s}=$ 91 GeV is studied.  The same photon identification
criteria are applied to the selected $\Zo\ra {\rm q \bar q}$ events.
An overall excess of ($20\pm10$)\% in the photon rate is found
in data with respect to the PYTHIA Monte Carlo which uses the
same JETSET fragmentation algorithm and particle decays as
KORALW.  A correction factor given by the ratio of photon production
rates in data and MC is determined as a function of the photon energy.
This correction is applied to the background component of ${\rm q \bar
  q}\gamma$ MC events as well as to the hadronic side of the
$\Wp\Wm\gamma$ MC events.  The uncertainty on this correction is
propagated to the measurement as a systematic uncertainty on the
$\Wp\Wm\gamma$ cross section which corresponds to $\pm$ 15 fb.\\

\section*{Determination of Anomalous Quartic Gauge Couplings}

The selected $\Wp\Wm\gamma$ events allow us to constrain anomalous
contributions to the SM quartic gauge boson vertices. In the framework
of References \citen{bud} and \citen{anja}, 
the extended Lagrangian includes new
dimension-6 operators, \bea
\label{L0}
{\cal L}_0 &=& - \frac{e^2}{16}\, \frac{a_0}{\Lambda^2}\, F^{\mu \nu}
\, F_{\mu
  \nu} \vector{W^{\alpha}} \cdot \vector{W_{\alpha}} \no \\
{\cal L}_c &=& - \frac{e^2}{16}\, \frac{a_c}{\Lambda^2}\, F^{\mu
  \alpha} \, F_{\mu
  \beta} \vector{W^{\beta}} \cdot \vector{W_{\alpha}} \no \\
{\cal L}_n &=& - \frac{i e^2}{16}\, \frac{a_n}{\Lambda^2}\,
\epsilon_{ijk} W_{\mu \alpha}^{(i)} W_{\nu}^{(j)} W^{(k)\alpha} F^{\mu
  \nu}, \no \ea where $a_0/\Lambda^2$, $a_c/\Lambda^2$, and
$a_n/\Lambda^2$ are the AQGCs, and $\Lambda$ represents the energy
scale for new physics.  The two parameters $a_0/\Lambda^2$ and
$a_c/\Lambda^2$, which are separately C and P conserving, generate
anomalous $\Wp\Wm\gamma\gamma$ and $\Zo\Zo\gamma\gamma$ vertices. The
term $a_n/\Lambda^2$, which is CP violating, gives rise to an
anomalous contribution to $\Wp\Wm\Zo\gamma$.  Although there are
already direct~\cite{zgg,opal-qgc} and indirect~\cite{indi} limits on
$a_0/\Lambda^2$ and $a_c/\Lambda^2$, only the study of $\Wp\Wm\gamma$
events allows for a direct measurement of the anomalous coupling
$a_n/\Lambda^2$ through the $\Wp\Wm \Zo\gamma$ vertex.

The EEWWG program implements the effects of the AQGCs through
the extended SM Lagrangian.  Figure~\ref{fig:xsec} shows how the
anomalous coupling $a_n/\Lambda^2$ manifests itself through a
deviation of the total cross section.

The anomalous component from the above operators is linear in the
photon energy at the matrix element level~\cite{anja}.  This implies
that also the shape of the photon spectrum is affected by AQGCs, in
particular, the hard part of the energy distribution (see
Figure~\ref{fig:sele}a).  The expected distribution for any value of
the three AQGCs is obtained by reweighting each KORALW MC event with
the ratio ${\cal W}(E_\gamma,a_0,a_c,a_n)$ of the known differential
distributions of E$_\gamma$ at generator level:
\bea {\cal W}(E_\gamma,a_0,a_c,a_n) ~ = ~ \frac{d{\sigma}^{\it
    EEWWG}}{dE_\gamma}(a_0,a_c,a_n) \left /~ \frac{d\sigma^{\it
      KORALW}}{dE_\gamma} \right. .\no \ea The reweighting procedure
is applied only to the ISR component of the MC selected sample, while
the FSR (from KORALW) and the background components of accepted events
are kept fixed.  The possible dependence of the selection efficiency
on the photon polar angle and on the angular separation from the
charged fermions in the event is found to be negligible.

Both the shape and the normalisation of the observed photon spectrum
in the range from 5 \GeV ~to 30 \GeV ~are used in a maximum-likelihood
fit to each of the anomalous couplings $a_0/\Lambda^2$,
$a_c/\Lambda^2$ and $a_n/\Lambda^2$, fixing the other two to zero.
The effects of the same systematic uncertainties described for the
cross section measurement are included, yielding the
68\% CL intervals: 
\begin{eqnarray*} 
-0.028 \GeV^{-2} & < & a_0/\Lambda^2 ~ < ~ 0.028 \GeV^{-2}\\ 
\phantom{0}-0.04 \GeV^{-2} & < & a_c/\Lambda^2 ~ < ~ 0.09\phantom{0} \GeV^{-2}\\ 
\phantom{0}-0.26 \GeV^{-2} & < & a_n/\Lambda^2 ~ < ~ 0.23\phantom{0} \GeV^{-2}.
\end{eqnarray*}
The results are in good agreement with the SM value of zero for each of
the anomalous quartic gauge couplings.
The 1-parameter limits at 95\% CL are:
\begin{eqnarray*} 
-0.045 \GeV^{-2} & < & a_0/\Lambda^2 ~ < ~ 0.045 \GeV^{-2}\\ 
\phantom{0}-0.08 \GeV^{-2} & < & a_c/\Lambda^2 ~ < ~ 0.13\phantom{0} \GeV^{-2}\\ 
\phantom{0}-0.41 \GeV^{-2} & < & a_n/\Lambda^2 ~ < ~ 0.37\phantom{0} \GeV^{-2}.
\end{eqnarray*}

\section*{The \boldmath$\epem \ra \nu\bar\nu\g\g$ Process }

The selection of acoplanar multi-photon events is identical to that
described in Reference \citen{papl3gg}.  At least two photons with
energies greater than 5 \GeV ~and 1 \GeV ~are required, with polar
angles between 14$^\circ$ and 166$^\circ$.  The
KORALZ~\cite{koralz_new} and NUNUGPV~\cite{NUNUGPV_new} Monte Carlo
generators are used to model the $\epem \ra \nu\bar\nu\g\g$ process
according to the SM.  The effects of the AQGCs \ao\ and \ac\ are
simulated using the EENUNUGGANO program~\cite{stir_phot}.  Note that
$\nu\bar\nu\g\g$ production is not sensitive to the
$a_n/\Lambda^2$~coupling.

We select 14 events at $\rts = 183 \GeV$ and 21 events at $\rts = 189
\GeV$ compared to a SM expectation of 13.3 and 36.2 events
respectively.

The EENUNUGGANO program does not describe the effects of the SM
$s$-channel Z exchange diagrams and the interference between these
diagrams and the $\Wp\Wm$ fusion diagram containing the $\Wp\Wm\g\g$
vertex.  Therefore additional cuts are applied to suppress the SM
contribution.  The energy of both photons must be greater than 10~\GeV.
If both photons are in the barrel region ($|\cos\theta|<0.7$), either
the recoil mass must be less than 80~\GeV\ or the sum of the photon
energies must be greater than 100~\GeV.  If one or two photons are in
the endcaps, where the SM contribution is larger, the recoil mass must
be less than 75~\GeV.  After applying these cuts no data event is
selected, consistent with the SM expectation of 0.15 events.

The expected number of events for any AQGC value is calculated based
on a sample of ten thousand simulated EENUNUGGANO events generated for
several values of \ao\ and \ac.  Its matrix element is used to
reweight the events to any AQGC value required, testing the procedure
by comparing the reweighted distributions to those from
samples generated at various values of AQGCs.  In all cases good
agreement is observed.

Since the program does not include higher order
corrections due to ISR, these effects are estimated by implementing
the EXCALIBUR~\cite{exca} collinear radiator function.  The cross
section is reduced by about 16\% which is used in the following. The
remaining theoretical uncertainty of 5\%~\cite{priv} is taken into
account in the AQGC limits.  The systematic uncertainty on the
selection efficiency~\cite{papl3gg} gives a much smaller contribution.

The 95\% CL upper limit on the number of expected events from the AQGC
signal is obtained taking into account the systematic error on the
accepted cross section; this corresponds to the following 1-parameter
limits at 95\% CL:
\begin{eqnarray*} 
-0.067 \GeV^{-2} & < & a_0/\Lambda^2 ~ < ~ 0.066 \GeV^{-2}\\ 
\phantom{0}-0.18 \GeV^{-2} & < & a_c/\Lambda^2 ~ < ~ 0.18\phantom{0}  \GeV^{-2}\,.
\end{eqnarray*}

\section*{Conclusion}

All results obtained results are in a good agreement with the SM
expectation of zero for each anomalous quartic gauge boson couplings.

Combining the results on \ao\ and \ac\ from our analyses of
$\Wp\Wm\gamma$ and $\nu\bar\nu \g\g$ production, we derive the
following 1-parameter 95\% CL limits:
\begin{eqnarray*} 
-0.043 \GeV^{-2} & < & a_0/\Lambda^2 ~ < ~ 0.043 \GeV^{-2}\\ 
\phantom{0}-0.08 \GeV^{-2} & < & a_c/\Lambda^2 ~ < ~ 0.13\phantom{0} \GeV^{-2}\\
\phantom{0}-0.41 \GeV^{-2} & < & a_n/\Lambda^2 ~ < ~ 0.37\phantom{0} \GeV^{-2}.
\end{eqnarray*}

\section*{Acknowledgements}

We thank J. W. Stirling and A. Werthenbach for providing us with the
$\epem\ra\Wp\Wm\gamma$ and the $\epem\ra\nu\bar\nu\gamma\gamma$
analytical calculations including AQGCs.  We wish to express our
gratitude to the CERN accelerator divisions for the superb performance
and the continuous and successful upgrade of the LEP machine.  We
acknowledge the contributions of the engineers and technicians who
have participated in the construction and maintenance of this
experiment.

%
\newpage
\typeout{   }     
\typeout{Using author list for paper 215 ONLY! }
\typeout{$Modified: Tue Jul 18 08:23:56 2000 by clare $}
\typeout{!!!!  This should only be used with document option a4p!!!!}
\typeout{   }
%
%
%
%
%
%

\newcount\tutecount  \tutecount=0
\def\tutenum#1{\global\advance\tutecount by 1 \xdef#1{\the\tutecount}}
\def\tute#1{$^{#1}$}
\tutenum\aachen            
\tutenum\nikhef            
\tutenum\mich              
\tutenum\lapp              
\tutenum\basel             
\tutenum\lsu               
\tutenum\beijing           
\tutenum\berlin            
\tutenum\bologna           
\tutenum\tata              
\tutenum\ne                
\tutenum\bucharest         
\tutenum\budapest          
\tutenum\mit               
\tutenum\debrecen          
\tutenum\florence          
\tutenum\cern              
\tutenum\wl                
\tutenum\geneva            
\tutenum\hefei             
\tutenum\seft              
\tutenum\lausanne          
\tutenum\lecce             
\tutenum\lyon              
\tutenum\madrid            
\tutenum\milan             
\tutenum\moscow            
\tutenum\naples            
\tutenum\cyprus            
\tutenum\nymegen           
\tutenum\caltech           
\tutenum\perugia           
\tutenum\cmu               
\tutenum\prince            
\tutenum\rome              
\tutenum\peters            
\tutenum\potenza           
\tutenum\salerno           
\tutenum\ucsd              
\tutenum\santiago          
\tutenum\sofia             
\tutenum\korea             
\tutenum\alabama           
\tutenum\utrecht           
\tutenum\purdue            
\tutenum\psinst            
\tutenum\zeuthen           
\tutenum\eth               
\tutenum\hamburg           
\tutenum\taiwan            
\tutenum\tsinghua          

{
\parskip=0pt
\noindent
{\bf The L3 Collaboration:}
\ifx\selectfont\undefined
 \baselineskip=10.8pt
 \baselineskip\baselinestretch\baselineskip
 \normalbaselineskip\baselineskip
 \ixpt
\else
 \fontsize{9}{10.8pt}\selectfont
\fi
\medskip
\tolerance=10000
\hbadness=5000
\raggedright
\hsize=162truemm\hoffset=0mm
\def\r{\rlap,}
\noindent

M.Acciarri\r\tute\milan\
P.Achard\r\tute\geneva\ 
O.Adriani\r\tute{\florence}\ 
M.Aguilar-Benitez\r\tute\madrid\ 
J.Alcaraz\r\tute\madrid\ 
G.Alemanni\r\tute\lausanne\
J.Allaby\r\tute\cern\
A.Aloisio\r\tute\naples\ 
M.G.Alviggi\r\tute\naples\
G.Ambrosi\r\tute\geneva\
% ...
\newpage
\begin{list}{A}{\itemsep=0pt plus 0pt minus 0pt\parsep=0pt plus 0pt minus 0pt
                \topsep=0pt plus 0pt minus 0pt}
\item[\aachen]
 I. Physikalisches Institut, RWTH, D-52056 Aachen, FRG$^{\S}$\\
 III. Physikalisches Institut, RWTH, D-52056 Aachen, FRG$^{\S}$
\item[\nikhef] National Institute for High Energy Physics, NIKHEF, 
     and University of Amsterdam, NL-1009 DB Amsterdam, The Netherlands
\item[\mich] University of Michigan, Ann Arbor, MI 48109, USA
\item[\lapp] Laboratoire d'Annecy-le-Vieux de Physique des Particules, 
     LAPP,IN2P3-CNRS, BP 110, F-74941 Annecy-le-Vieux CEDEX, France
\item[\basel] Institute of Physics, University of Basel, CH-4056 Basel,
     Switzerland
\item[\lsu] Louisiana State University, Baton Rouge, LA 70803, USA
\item[\beijing] Institute of High Energy Physics, IHEP, 
  100039 Beijing, China$^{\triangle}$ 
\item[\berlin] Humboldt University, D-10099 Berlin, FRG$^{\S}$
\item[\bologna] University of Bologna and INFN-Sezione di Bologna, 
     I-40126 Bologna, Italy
\item[\tata] Tata Institute of Fundamental Research, Bombay 400 005, India
\item[\ne] Northeastern University, Boston, MA 02115, USA
\item[\bucharest] Institute of Atomic Physics and University of Bucharest,
     R-76900 Bucharest, Romania
\item[\budapest] Central Research Institute for Physics of the 
     Hungarian Academy of Sciences, H-1525 Budapest 114, Hungary$^{\ddag}$
\item[\mit] Massachusetts Institute of Technology, Cambridge, MA 02139, USA
\item[\debrecen] KLTE-ATOMKI, H-4010 Debrecen, Hungary$^\P$
\item[\florence] INFN Sezione di Firenze and University of Florence, 
     I-50125 Florence, Italy
\item[\cern] European Laboratory for Particle Physics, CERN, 
     CH-1211 Geneva 23, Switzerland
\item[\wl] World Laboratory, FBLJA  Project, CH-1211 Geneva 23, Switzerland
\item[\geneva] University of Geneva, CH-1211 Geneva 4, Switzerland
\item[\hefei] Chinese University of Science and Technology, USTC,
      Hefei, Anhui 230 029, China$^{\triangle}$
\item[\seft] SEFT, Research Institute for High Energy Physics, P.O. Box 9,
      SF-00014 Helsinki, Finland
\item[\lausanne] University of Lausanne, CH-1015 Lausanne, Switzerland
\item[\lecce] INFN-Sezione di Lecce and Universit\'a Degli Studi di Lecce,
     I-73100 Lecce, Italy
\item[\lyon] Institut de Physique Nucl\'eaire de Lyon, 
     IN2P3-CNRS,Universit\'e Claude Bernard, 
     F-69622 Villeurbanne, France
\item[\madrid] Centro de Investigaciones Energ{\'e}ticas, 
     Medioambientales y Tecnolog{\'\i}cas, CIEMAT, E-28040 Madrid,
     Spain${\flat}$ 
\item[\milan] INFN-Sezione di Milano, I-20133 Milan, Italy
\item[\moscow] Institute of Theoretical and Experimental Physics, ITEP, 
     Moscow, Russia
\item[\naples] INFN-Sezione di Napoli and University of Naples, 
     I-80125 Naples, Italy
\item[\cyprus] Department of Natural Sciences, University of Cyprus,
     Nicosia, Cyprus
\item[\nymegen] University of Nijmegen and NIKHEF, 
     NL-6525 ED Nijmegen, The Netherlands
\item[\caltech] California Institute of Technology, Pasadena, CA 91125, USA
\item[\perugia] INFN-Sezione di Perugia and Universit\'a Degli 
     Studi di Perugia, I-06100 Perugia, Italy   
\item[\cmu] Carnegie Mellon University, Pittsburgh, PA 15213, USA
\item[\prince] Princeton University, Princeton, NJ 08544, USA
\item[\rome] INFN-Sezione di Roma and University of Rome, ``La Sapienza",
     I-00185 Rome, Italy
\item[\peters] Nuclear Physics Institute, St. Petersburg, Russia
\item[\potenza] INFN-Sezione di Napoli and University of Potenza, 
     I-85100 Potenza, Italy
\item[\salerno] University and INFN, Salerno, I-84100 Salerno, Italy
\item[\ucsd] University of California, San Diego, CA 92093, USA
\item[\santiago] Dept. de Fisica de Particulas Elementales, Univ. de Santiago,
     E-15706 Santiago de Compostela, Spain
\item[\sofia] Bulgarian Academy of Sciences, Central Lab.~of 
     Mechatronics and Instrumentation, BU-1113 Sofia, Bulgaria
\item[\korea]  Laboratory of High Energy Physics, 
     Kyungpook National University, 702-701 Taegu, Republic of Korea
\item[\alabama] University of Alabama, Tuscaloosa, AL 35486, USA
\item[\utrecht] Utrecht University and NIKHEF, NL-3584 CB Utrecht, 
     The Netherlands
\item[\purdue] Purdue University, West Lafayette, IN 47907, USA
\item[\psinst] Paul Scherrer Institut, PSI, CH-5232 Villigen, Switzerland
\item[\zeuthen] DESY, D-15738 Zeuthen, 
     FRG
\item[\eth] Eidgen\"ossische Technische Hochschule, ETH Z\"urich,
     CH-8093 Z\"urich, Switzerland
\item[\hamburg] University of Hamburg, D-22761 Hamburg, FRG
\item[\taiwan] National Central University, Chung-Li, Taiwan, China
\item[\tsinghua] Department of Physics, National Tsing Hua University,
      Taiwan, China
\item[\S]  Supported by the German Bundesministerium 
        f\"ur Bildung, Wissenschaft, Forschung und Technologie
\item[\ddag] Supported by the Hungarian OTKA fund under contract
numbers T019181, F023259 and T024011.
\item[\P] Also supported by the Hungarian OTKA fund under contract
  numbers T22238 and T026178.
\item[$\flat$] Supported also by the Comisi\'on Interministerial de Ciencia y 
        Tecnolog{\'\i}a.
\item[$\sharp$] Also supported by CONICET and Universidad Nacional de La Plata,
        CC 67, 1900 La Plata, Argentina.
\item[$\diamondsuit$] Also supported by Panjab University, Chandigarh-160014, 
        India.
\item[$\triangle$] Supported by the National Natural Science
  Foundation of China.
\end{list}
}
\vfill


\newpage
%
\bibliographystyle{l3stylem}

\begin{mcbibliography}{10}

\bibitem{sm_glashow}
S.L. Glashow,
\newblock  Nucl. Phys. {\bf 22}  (1961) 579\relax
\relax
\bibitem{sm_salam}
A. Salam,
\newblock  in Elementary Particle Theory, ed. {N.~Svartholm},  (Alm\-qvist and
  Wiksell, Stockholm, 1968), p. 367\relax
\relax
\bibitem{sm_weinberg}
S. Weinberg,
\newblock  Phys. Rev. Lett. {\bf 19}  (1967) 1264\relax
\relax
\bibitem{velt}
M. Veltman, Nucl. Phys. {\bf B7} (1968) 637; G.M.'t Hooft, Nucl. Phys. {\bf
  B35} (1971) 167; G.M.'t Hooft and M. Veltman, Nucl. Phys. {\bf B44} (1972)
  189; Nucl. Phys. {\bf B50} (1972) 318\relax
\relax
\bibitem{stir_phot}
J.W.~Stirling and A.~Werthenbach,
\newblock  Phys. Lett. {\bf B466}  (1999) 369\relax
\relax
\bibitem{l3det}
L3 Collab., B.~Adeva $\etal$, Nucl. Instr. Meth. {\bf A289} (1990) 35;
  J.~A.~Bakken $\etal$, Nucl. Instr. Meth. {\bf A275} (1989) 81; O.~Adriani
  $\etal$, Nucl. Instr. Meth. {\bf A302} (1991) 53; B.~Adeva $\etal$, Nucl.
  Instr. Meth. {\bf A323} (1992) 109; K.~Deiters $\etal$, Nucl. Instr. Meth.
  {\bf A323} (1992) 162; M.~Chemarin $\etal$, Nucl. Instr. Meth. {\bf A349}
  (1994) 345; M.~Acciarri $\etal$, Nucl. Instr. Meth. {\bf A351} (1994) 300;
  G.~Basti $\etal$, Nucl. Instr. Meth. {\bf A374} (1996) 293; A.~Adam $\etal$,
  Nucl. Instr. Meth. {\bf A383} (1996) 342\relax
\relax
\bibitem{bud}
G. Belanger and F. Boudjema,
\newblock  Nucl. Phys. {\bf B 288}  (1992) 201\relax
\relax
\bibitem{anja}
J.W. Stirling and A. Werthenbach, \EPJ {\bf C14} (2000) 103\relax
\relax
\bibitem{opal-qgc}
Opal Collab., G. Abbiendi {\em et al.},
\newblock  Phys. Lett. {\bf B471}  (1999) 293\relax
\relax
\bibitem{zgg}
L3 \coll, M. Acciarri \etal, \PL {\bf B478} (2000) 34\relax
\relax
\bibitem{wk}
KORALW V1.42, M. Skrzypek {\it et al.} {Comp. Phys. Comm.} {\bf 94} (1996) 216;
  Phys. Lett. {\bf B372} (1996) 289\relax
\relax
\bibitem{photos}
E. Barberio, B. van Eijk and Z. Was, { Comp. Phys. Comm.} {\bf 79} (1994)
  291\relax
\relax
\bibitem{js}
T. Sj{\"o}strand and H. U. Bengtsson, {Comp. Phys. Comm.} {\bf 46} (1987)
  43\relax
\relax
\bibitem{pyt}
T. Sj{\"o}strand, {Comp. Phys. Comm.} {\bf 82} (1994) 74\relax
\relax
\bibitem{geant}
The L3 detector simulation is based on GEANT Version 3.15.\\ See R. Brun \etal,
  ``GEANT 3'', CERN DD/EE/84-1 (Revised), September 1987.\\ The GHEISHA program
  (H. Fesefeldt, RWTH Aachen Report PITHA 85/02 (1985)) is used to simulate
  hadronic interactions.\relax
\relax
\bibitem{priv}
J.W. Stirling and A. Werthenbach, private communication\relax
\relax
\bibitem{exca}
F.A. Berends, R. Pittau and R. Kleiss {Comp. Phys. Comm.} {\bf 85} (1995)
  437\relax
\relax
\bibitem{yfs}
S. Jadach, {\it et al.}, preprint UTHEP-00-0101, to appear\relax
\relax
\bibitem{wkpeople}
S. Jadach, private communication\relax
\relax
\bibitem{l3-pre98}
L3 Collab., M.~Acciarri {\em et al.}, Preprint CERN-EP-2000-104 (2000)\relax
\relax
\bibitem{indi}
O.J.P. Eboli and M.C. Gonzales--Garcia, Phys. Lett. {\bf B411} (1994) 381\relax
\relax
\bibitem{papl3gg}
L3 \coll, M. Acciarri \etal, \PL {\bf B444} (1998) 503, {\bf B470} (1999)
  268\relax
\relax
\bibitem{koralz_new}
The KORALZ version 4.03 is used.\\ S. Jadach, B.F.L. Ward and Z. Was, \CPC {\bf
  79} (1994) 503\relax
\relax
\bibitem{NUNUGPV_new}
G. Montagna, {\em et al.},
\newblock  Nucl. Phys. {\bf B541}  (1999) 31\relax
\relax
\end{mcbibliography}

            
\clearpage
            
\vspace*{8cm}

\input{feynman}\bigphotons      

\end{document}